\documentclass{article}

\usepackage[table]{xcolor}
\usepackage{mathtools}
\usepackage{array}
\usepackage{float}
\usepackage{easy-todo}
\usepackage[none]{hyphenat}
\usepackage{flafter}
\usepackage{placeins}
\usepackage{times}
\usepackage{cite}
\usepackage{amsmath}
\usepackage{latexsym}
\usepackage{pstricks}
\usepackage{setspace}
\usepackage{siunitx}
\usepackage{fancyhdr}
\usepackage{array}
\usepackage{float}
\usepackage{setspace}

\usepackage{graphicx}

\usepackage{acronym}
\usepackage{url}
\usepackage{easy-todo}
\usepackage{subfig}
\usepackage{multirow}
\usepackage[left=1.25in,right=1.25in,top=1in,bottom=1in]{geometry} 
\usepackage{algorithm}
\usepackage{algorithmic}
\pdfoutput=1
\title{GSM--CommSense-based through-the-wall sensing}

\author{\textbf{Abhishek Bhatta}\\
Electrical Engineering Department\\
University of Cape Town
\\
Email: bhattaomatic@gmail.com\\
\and
\textbf{Amit Kumar Mishra}\\
Electrical Engineering Department\\
University of Cape Town
\\
Email: akmishra@ieee.org}

\begin{document}

\maketitle
\begin{abstract}
We have recently proposed a scheme to use the channel equalization blocks of telecommunication systems to sense changes in an environment. 
We call this communication-sensing, CommSense for short. 
After some initial positive results we tried to use our global system for mobile communication (GSM) based CommSense system for a through-the-wall sensing application.  
As the system was inherently highly under-determined we used statistical machine learning techniques to help us sense environmental changes in the behind-the-wall experiments. We observed that with limited amount of data per GSM frame of $577$~\si{\micro\second} a person can be detected across a wall to an accuracy of $77.458\%$ and a person carrying a weapon can be detected to an accuracy of $95.208\%$.
We present details of the experiments and the encouraging results that we have obtained in this article. 
\end{abstract}


\section{Introduction}\label{sec:intro}

 Combining radar and telecommunication system capabilities has been a major research focus for a long time \cite{sym_pat}. This includes systems like passive or commensal radars \cite{thomas2006ambiguity, inggs2014planning} that use the already available radio-frequency (RF) broadcast transmitters to detect and classify target parameters such as location, velocity, etc..  
 
\begin{figure}[t]
\centering
\includegraphics[width=0.8\textwidth]{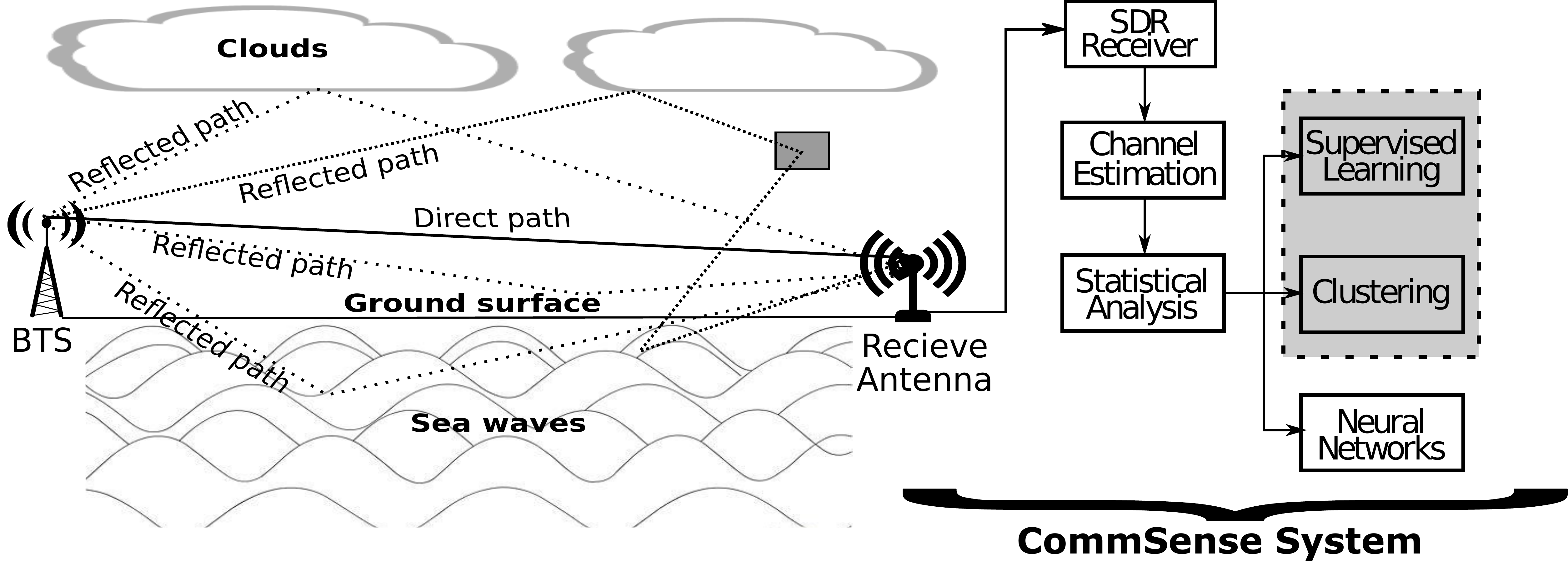}
	\caption{Concept diagram showing multipath signals from transmitter to receiver and the blocks used in the GSM--CommSense system (the shaded region is the focus of the current work). The base transceiver station (BTS) transmits the signal.}\label{fig:conceptdiag}
\end{figure}

 In our work we started with the initial hypothesis that the channel estimation block of telecommunication systems can be used to gather information about the immediate environment \cite{mishra2015patent}, a system which we call communication-sensing (CommSense). The basic concept diagram of the CommSense system is shown in Figure \ref{fig:conceptdiag}. In the previous work \cite{bhatta2017gsm}, we used this system to detect various environmental parameters and the results were promising.
Thus we decided to explore additional application of this system. In this article we demonstrate the ability of CommSense to use global system for mobile communication (GSM) based signal to achieve through-the-wall sensing. From here on we will refer to this system as GSM--CommSense.
 
 The current literature on using radar systems for through-the-wall sensing relies on either a system that uses a dedicated active transmitter \cite{li2012through}, or a passive system without a dedicated active transmitter that requires a dedicated receiver pointed at the reference with another pointed at the scene (surveillance)\cite{chetty2012through}. The reference and the surveillance channel information is then correlated to extract the target information. The GSM--CommSense system is designed to eliminate the need of the reference antenna and use the channel estimation techniques of GSM systems to extract the channel information.
 One of the major challenges of the GSM--CommSense system is the lack of sufficient amounts of data. 
For example in the initial design, GSM--CommSense only outputted $40$ samples per data frame received. This was increased to $48$ samples by revising the design and adjusting the buffer size which is limited by the hardware capabilities. 
 Getting a proper phenomenological sense from this amount of data did not seem possible. 
 Hence, we used the application specific instrumentation framework (ASIN) \cite{mishra_10_asin, sardar_14} which uses statistical machine learning to recognize `events of interest'. 
 The shaded region in Figure \ref{fig:conceptdiag} is the major focus of the work presented here, which corresponds to the analysis of the data captured by the GSM--CommSense system.
 We show some encouraging results obtained from the limited amount of experiments we have conducted to sense objects and events behind the wall. 
 
 

\section{GSM--CommSense: A Short Review}\label{sec:commsense}
 A channel, also known as a physical transmission medium, introduces interference to the signal being transmitted through it. These interferences are proportional to the properties of the channel itself. In communication systems this type of interference causes degradation of the information and thus various channel equalization techniques are used to minimize the channel effects \cite{pukkila2000channel,bhatta2015implementation,newson1991adaptive}. In case of a GSM system the transmitter transmits a known bit stream referred to as a training sequence which is used at the receiver to equalize the effects of the channel. The GSM--CommSense system \cite{bhatta2017gsm,bhatta2016gsmconf} is designed to use the training sequence in order to extract the channel impulse response (CIR) and classify different environmental scenarios. The current work specifically focuses on sensing through-the-wall.
 
 The GSM--CommSense system \cite{mishra2015patent}, although based on the idea of a radar systems, is not a typical radar system as it does not process the information using correlation between a reference signal and surveillance signal. Instead it uses the training sequence to extract the channel effects. This eliminates the need for a reference channel which aids in reducing the overall cost of the design and makes it smaller, more portable system. One of the disadvantages of using the channel equalization to extract the channel parameters is that it reduces the amount of data available for identifying events of interest, as only $10-15\%$ of the information transmitted over the channel is used for equalization. 
 
 In \cite{bhatta2017gsm} the basic implementation of GSM--CommSense is presented and the received data is statistically analysed for different environmental conditions. The system design and hardware considerations are discussed in detail in \cite{bhatta2016gsmconf}. The major novelties of the GSM--CommSense system include using the training sequence to estimate the channel, eliminating the need for a reference antenna, all while being implemented on an open source, hardware-software platform, keeping it open for future modifications.

\section{Experimental Setup}\label{sec:expsetup}
Terminologies used in the rest of the article. 
\begin{itemize}
\item Event: An event is defined as a general test set-up in which data is captured and analysed.
\item Test Scenario: A particular set-up under an event is defined here as a test scenario. Each event has multiple test scenarios.
\item Set: A set represents data captured for $30$ \si{\second}, which is used for analysis. Each test scenario has multiple sets.
\item Case: A case is used to differentiate between the particular set used for testing. Each case uses one set as test data and others as training data.
\end{itemize}

 The objective in the current work is to demonstrate the ability of the GSM--CommSense system to sense events across a brick wall of thickness $222$ \si{\milli\meter}. There are in total two distinct events analysed in this work as described below. 
\begin{itemize}
\item[(a)] Presence (in motion/stationary) and absence of a person in `Room 2' with the receiver placed in `Room 1'
\item[(b)] Presence of two persons in `Room 2' one of whom is carrying a weapon with the receiver placed in `Room 1'.
\end{itemize}

\begin{figure}
\centering
\includegraphics[width=0.4\textwidth]{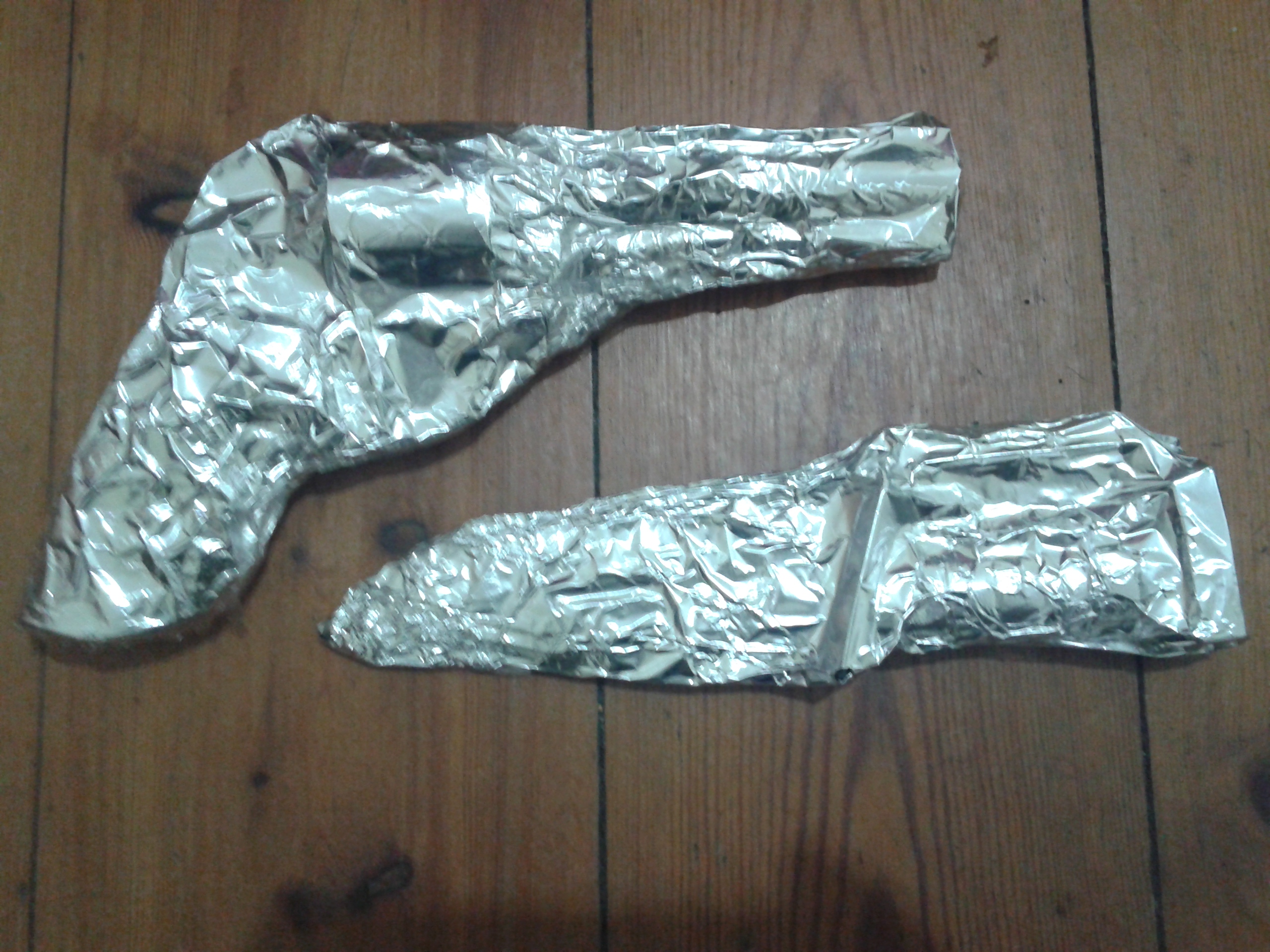}
	\caption{Plastic toy weapons covered with aluminium foil to increase the reflectivity is used for testing.}\label{fig:weapons}
\end{figure}

 The test is performed in a house with the layout as shown in Figure \ref{fig:setuplayout}. The separations show different rooms in the house and the dotted lines are the location of the doors. The weapons used for the purpose of this test are toy gun and a knife covered with aluminium foil to increase the radar cross section (RCS) (shown in Figure \ref{fig:weapons}). The aluminium foil is placed as the toy weapons are made of plastic and in reality most of the concealed weapons are made of metal. All the datasets of event (a) were captured within the duration of an hour keeping the external parameters such as location of the objects within the room, temperature of the room, etc., constant for the duration of the test. The experiments for event (b) were conducted on a different day from event (a). This test was also performed within the duration of an hour and the external parameters, as mentioned above, were kept constant for the duration of the test.
 The receiver is placed in `Room 1' at a distance of $1$ \si{\meter} from the wall and at a height of $1.12$ \si{\meter} from the floor.
The different test scenarios are described below.

\subsection{Scenarios for Event (a)}
\subsubsection{Person stationary}
 Person standing in `Room 2' at a distance of $3$ \si{\meter} from the wall. Three sets of data are captured and analysed for this scenario. Each set contains data received and sampled for a $30$ \si{\second} interval with a break of $20$ \si{\second} in between consecutive sets. This time interval was chosen to create statistical rigour to the received data.
 
\subsubsection{No Person}
 In this scenario there is no person in `Room 2' maintaining all the other parameters of the test. This scenario also contains the same amount of datasets with the same duration as the previous case.
 
\subsubsection{Person Moving}
\begin{figure}
\centering
\includegraphics[width=0.5\textwidth]{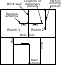}
	\caption{Layout of the experimental set-up.}\label{fig:setuplayout}
\end{figure}

 In this scenario the person is walking in a loop in `Room 2' at a distance of approximately $3$ \si{\meter} from the wall, as shown in Figure \ref{fig:setuplayout} for the entire duration of the test. This scenario also contains same amount of datasets as in the other two test cases.
 
\subsection{Scenarios for Event (b)}
\subsubsection{Concealed Weapon}
 There are two persons in `Room 2', one of them concealing a weapon under the jacket and the other is empty handed. Location of the person concealing the weapon is $3$ \si{\meter} from the brick wall and the location of the second person is randomized in the room, keeping their random position fixed for each capture. Each set is captured in the same way as the person-stationary scenario defined above. 

\subsubsection{No Weapon}
 In this scenario there are two persons in `Room 2' neither of them carrying a weapon. Location of person-$1$ is still at $3$ \si{\meter} from the wall and person-$2$ is randomly located in the room remaining stationary for the duration of the capture. 

\subsubsection{Visible Weapon}
 This scenario is similar to the concealed-weapon scenario with the only difference being that the weapons are visible and not concealed under jacket.

 
%
%

\section{Data Analysis}\label{sec:dataanalysis}
 Real time data is captured using the GSM--CommSense system for each of the scenarios defined in Section \ref{sec:expsetup}. The wireless GSM signal is received and digitally sampled using an analog-to-digital converter. The channel information is extracted from these samples and saved into a binary file for further analysis.
 The analysis is performed in two steps, first the data is passed through a clustering algorithm called Principal Component Analysis (PCA) \cite{abdi2010principal} to check if it contains any visually separable information. The limitations of this method is that we can only visually separate the information in three dimensions. This limits the percentage of information available for observation. Therefore, a supervised learning algorithm, Support Vector Machine (SVM) \cite{vapnik2013nature, chapelle2002choosing} is used to predict each of the cases based on learning from the others. The major limitation of SVM is that it can only predict an outcome for an event it has been trained on, so for each new scenario the algorithm has to be trained separately.
 Each of these analysis is performed directly on the dataset captured by the GSM--CommSense system.

\subsection{Principal Component Analysis (PCA)}
\subsubsection{Description}
 PCA is a statistical analysis tool which is used to transform an observation of possibly correlated variables into linearly uncorrelated components, called principal components. PCA is designed to provide the largest possible variance in the first principal component and each of the succeeding components has the highest possible variance with the constraint that it is orthogonal to the preceding component. This, when plotted provides the best possible visualization of the captured data. 
 
 In order to derive the principal components from a data first we need to generate the Singular Value Decomposition (SVD)\cite{abdi2007singular,abdi2007eigen} equivalent of the dataset. The SVD of the dataset represented by $\mathbf{A}$ is given as:

\begin{equation}\label{eq:svd}
\mathbf{A = U\Delta V}^\text{T} \quad .
\end{equation}

 Here $\mathbf{U}$ is a matrix of left singular vectors, $\mathbf{V}$ is a matrix of right singular values, and $\mathbf{\Delta}$ is a diagonal matrix of singular values. The principal component matrix $\mathbf{F}$, as derived from the data representation in Equation~(\ref{eq:svd}) is given as:

\begin{equation}\label{eq:principalComponents}
\mathbf{F = U\Delta} \quad .
\end{equation}

 Detailed calculations providing proper explanation of PCA and its implementation on the CommSense system is shown in \cite{bhatta2017gsm}.

\subsubsection{Results and Discussion}
\begin{figure}
\centering
\includegraphics[width=0.5\textwidth, height=0.3\textwidth]{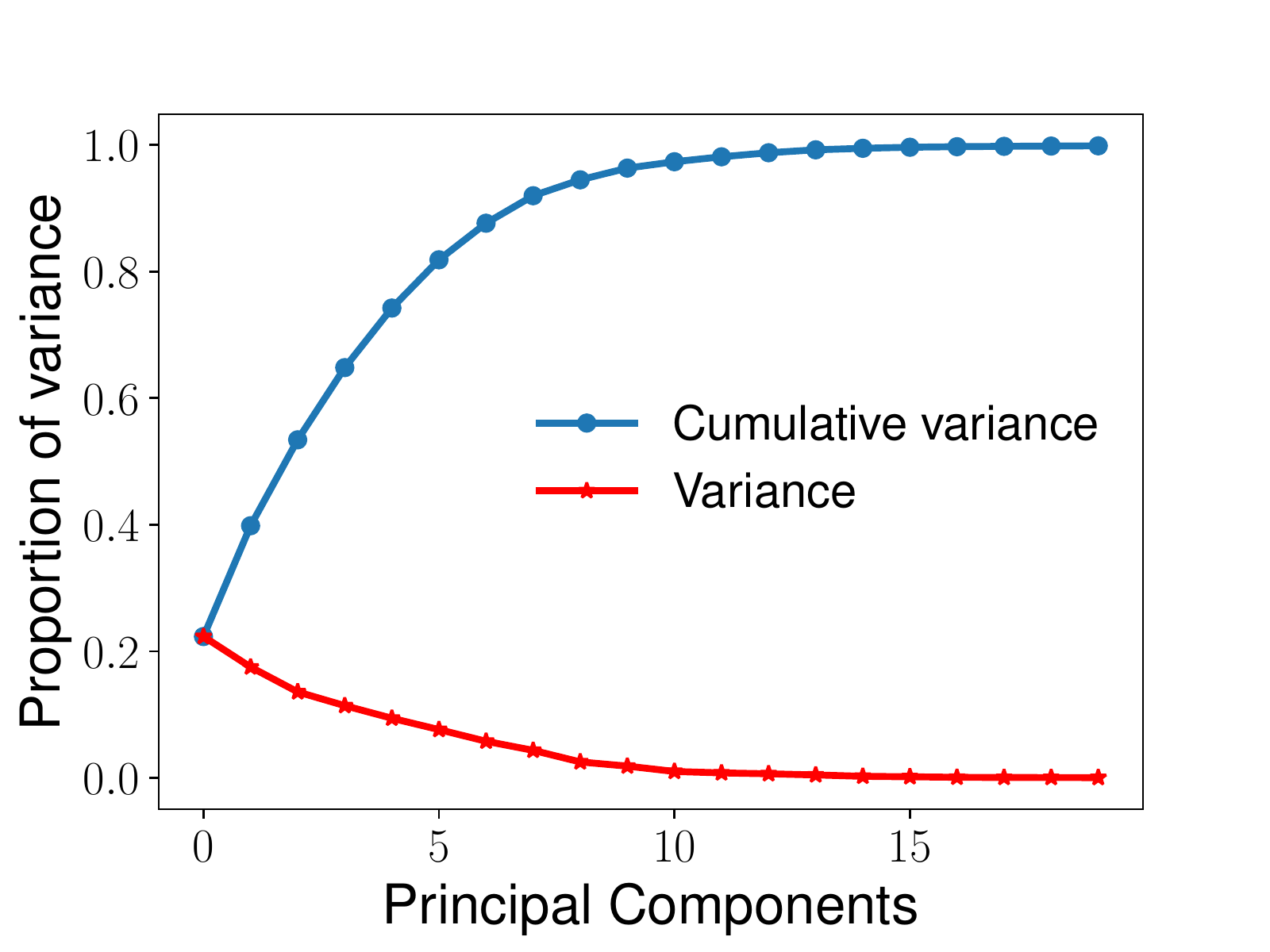}
	\caption{Scree plot showing the contribution of each principal component towards the total information content in the data.}\label{fig:pcascree}
\end{figure}

 Figure \ref{fig:pcascree} contains a plot showing the contribution of individual principal components towards the total information present in one of the recorded dataset. It can be observed that the first $12$ principal components constitute approximately $98\%$ of the information, based on second order moment, and the contribution of each of the components after that are negligible (less than $0.2\%$ per component). 
 
\begin{figure}
\centering
\subfloat[]{%
\resizebox*{0.43\textwidth}{!}{\includegraphics{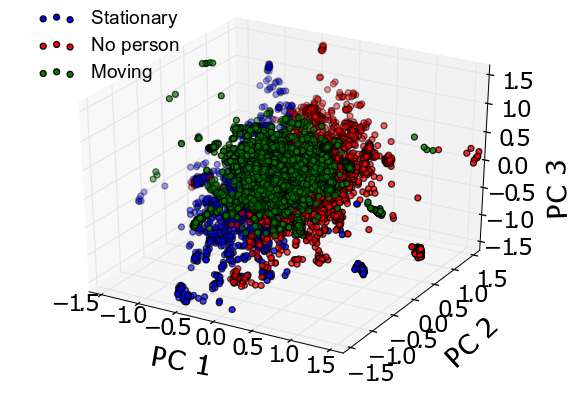}}\label{fig:pca3d}}\hspace{5pt}
\subfloat[]{%
\resizebox*{0.43\textwidth}{!}{\includegraphics{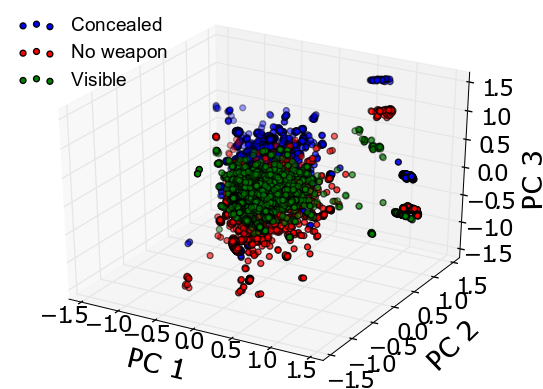}}\label{fig:pca3dweapon}}
\caption{PCA plots showing the clustering of through-the-wall data when the receiver is in `Room 1' and the event occurs in `Room 2'. The events are: \protect\subref{fig:pca3d} PCA clusters in the presence of a stationary person, a moving person and no person \protect\subref{fig:pca3dweapon} PCA clusters of a person with a concealed weapon, the person without a weapon and the person carrying a visible weapon.}
\end{figure}

 The PCA plots for the first three components of each dataset for event (a) are shown in Figure \ref{fig:pca3d} and event (b) are shown in Figure \ref{fig:pca3dweapon}. Although the first three components only contain about $52\%$ of the total information, some differences in the clusters are visible. This is well within the range of expected outcomes because the test is to look for the change in one event while keeping everything else constant. Each of the colors in the PCA plots show details of a particular test scenario. In Figure \ref{fig:pca3d} the blue color represents the condition when a person is stationary, the green color represents the condition when the person is moving within the room and the red color represents the condition when there is no person in `Room 2' with the receiver in `Room 1'. Similarly Figure \ref{fig:pca3dweapon} the blue color represents the condition when a person is concealing a weapon under a jacket, the green color represents the condition when the weapons are held visibly and the red color represents the condition when the person is not carrying any weapons in `Room 2' with the receiver in `Room 1'.
 
 To get these plots, only a part of set $1$ from each scenario is used which contains $3000\times40$ normalized data points for event (a) and $3000\times48$ normalized data points for event (b). This does not result in a clear separation of the data as most of the conditions are constant with the only change being the test scenario itself. Thus another analysis needs to be performed to understand the separation in the datasets from each of the scenarios. 

%

\subsection{Support Vector Machines (SVM)}
\subsubsection{Description}
 SVM is a supervised learning method that takes a labeled dataset, known as training data, and generates a model. This model can be used to categorize the test dataset. The objective of SVM is to design a hyperplane in high dimensional space that classifies training sets into different classes. There may exist multiple such hyperplanes that can achieve the same task of separating the classes, the best choice is the hyperplane that has maximum separation from nearest element from each class. In order to maximize the geometric margin between two classes and minimize the error, the soft margin SVM can be represented as:

\begin{equation}\label{eq:softmarginSVM}
\min_{\pmb{\alpha} \in R^{F}}\frac{1}{2} || \pmb{\alpha} ||_{2}^{2} + C\sum_{i=1}^{n}l(y_i,f_\alpha(x_i))	\quad .
\end{equation}

  Here, $\pmb{\alpha}$ is normal vector to the hyperplane that separates two classes of data, $l(.)$ is a loss function, $C$ is regularization parameter. $f_\alpha(x_i) =<\phi(x_i), \pmb{\alpha}>$, where, $\phi(x): R^d > R^F$ is a function that maps the training data points from input space, $R^d$ to a $F-$dimensional feature space, $R^F$. $||\cdot||_{2}$, represents the standard or the Euclidean norm. $< \cdot , \cdot>$, is the scalar product of two vectors. For large $F$ the inner products of the feature space can be calculated by a kernel function, $k(x,y) = \langle\phi(x),\phi(y)\rangle$. Detailed explanation regarding implementation of SVM on CommSense system is given in \cite{bhatta2017classification}.

\subsubsection{Results and Discussion}
 Details of the datasets used in SVM for the purpose of training and testing is given in Table \ref{tab:traintestdet}. After careful consideration and rigorous testing by varying the SVM parameters the penalty parameter is set to $1.0$, a linear kernel is used and, the influence of the training example given by, $\gamma$, is set to $0.025$ for event (a) and $0.0208$ for event (b) to yield the best results. In case of event (a) there are in total $39000\times40$ data points used for training and $12000\times40$ data points used for test. The training dataset is evenly distributed into the three scenarios defined above with $13000\times40$ points for each scenario. Since two sets are used for training, each set consists of $6500\times40$ points. In the case of event (b) the number of sample points from each frame changes from $40$ to $48$ thus in total $39000\times48$ data points used for training and $12000\times48$ data points used as test dataset. The additional $8$ points per dataset creates a noticeable difference in the prediction accuracy.
 The test dataset is evenly distributed among the three scenario and there are $4000\times40$ points in event (a) and $4000\times48$ points in event (b) per scenario. 
 
\begin{table}[tbph]
\caption{Description of the test set and training set to generate Table \ref{tab:svmpred} and \ref{tab:svmpredweapon} along with the number of data points used for training and testing.}\label{tab:traintestdet}
\centering
\begin{tabular}{ | c | c | c | c | c |} \hline
\multirow{2}{*}{\textbf{Case}} & \multicolumn{2}{c|}{\textbf{Training set}} & \multicolumn{2}{c|}{\textbf{Test Set}}\\\cline{2-5}
& Sets & No. of Data Points & Sets & No. of Data Points \\ \hline 
Case 1 & Set $2(a)$ and $3(a)$ & $39000\times40$ & Set $1(a)$ & $12000\times40$\\ \hline
Case 2 & Set $1(a)$ and $3(a)$ & $39000\times40$ & Set $2(a)$ & $12000\times40$\\ \hline
Case 3 & Set $1(a)$ and $2(a)$ & $39000\times40$ & Set $3(a)$ & $12000\times40$\\ \hline
Case 4 & Set $2(b)$ and $3(b)$ & $39000\times48$ & Set $1(b)$ & $12000\times48$\\ \hline
Case 5 & Set $1(b)$ and $3(b)$ & $39000\times48$ & Set $2(b)$ & $12000\times48$\\ \hline
Case 6 & Set $1(b)$ and $2(b)$ & $39000\times48$ & Set $3(b)$ & $12000\times48$\\ \hline
\end{tabular}
\end{table}

\begin{table}[tb]
\caption{Confusion matrix for detection of a person through a brick wall. All the values shown here are in percentages.}\label{tab:svmpred}
\resizebox{0.52\textwidth}{!}{\begin{minipage}{0.6\textwidth}
\centering
\begin{tabular}{ | >{\centering\arraybackslash}m{14mm} | >{\centering\arraybackslash}m{14mm} | >{\centering\arraybackslash}m{10mm} | >{\centering\arraybackslash}m{10mm} | >{\centering\arraybackslash}m{14mm} | >{\centering\arraybackslash}m{10mm} | >{\centering\arraybackslash}m{10mm} | >{\centering\arraybackslash}m{14mm} | >{\centering\arraybackslash}m{10mm} | >{\centering\arraybackslash}m{10mm} | >{\centering\arraybackslash}m{16mm} |} \hline
\textbf{Predicted Label} & \multicolumn{3}{c|}{\textbf{Case 1}} & \multicolumn{3}{c|}{\textbf{Case 2}} & \multicolumn{3}{c|}{\textbf{Case 3}} & \textbf{Average correct classification}\\ \hline
 \textbf{Person Stationary} & \cellcolor[gray]{.8}$86.000$ & $4.825$ & $9.175$ & \cellcolor[gray]{.8}$85.725$ & $ 3.525$ & $10.750$ & \cellcolor[gray]{.8}$86.125$ & $4.975$ & $8.900$ & $85.950$ \\ \hline
 \textbf{No Person} & $16.925$ & \cellcolor[gray]{.8}$71.925$ & $11.150$ & $8.250$ & \cellcolor[gray]{.8}$82.500$ & $9.250$ & $12.850$ & \cellcolor[gray]{.8}$77.950$ & $9.200$ & $77.458$ \\ \hline
 \textbf{Person Moving} & $9.775$ & $6.675$ & \cellcolor[gray]{.8}$83.550$ & $5.775$ & $4.275$ & \cellcolor[gray]{.8}$89.950$ & $10.550$ & $6.475$ & \cellcolor[gray]{.8}$82.975$ & $85.491$ \\ \hline
 & \textbf{Person Stationary} & \textbf{No Person} & \textbf{Person Moving} & \textbf{Person Stationary} & \textbf{No Person} & \textbf{Person Moving} & \textbf{Person Stationary} & \textbf{No Person} & \textbf{Person Moving}& \\ \hline
 & \multicolumn{9}{c|}{\textbf{True Label}} & \\ \hline
\end{tabular}
\end{minipage}}
\end{table} 

\begin{table}[tb]
\caption{Confusion matrix for detection of a person carrying weapon through a brick wall. All the values shown here are in percentages.}\label{tab:svmpredweapon}
\resizebox{0.52\textwidth}{!}{\begin{minipage}{0.6\textwidth}
\centering
\begin{tabular}{ | >{\centering\arraybackslash}m{14mm} | >{\centering\arraybackslash}m{14mm} | >{\centering\arraybackslash}m{10mm} | >{\centering\arraybackslash}m{10mm} | >{\centering\arraybackslash}m{14mm} | >{\centering\arraybackslash}m{10mm} | >{\centering\arraybackslash}m{10mm} | >{\centering\arraybackslash}m{14mm} | >{\centering\arraybackslash}m{10mm} | >{\centering\arraybackslash}m{10mm} | >{\centering\arraybackslash}m{16mm} |} \hline
\textbf{Predicted Label}  & \multicolumn{3}{c|}{\textbf{Case 4}} & \multicolumn{3}{c|}{\textbf{Case 5}} & \multicolumn{3}{c|}{\textbf{Case 6}} & \textbf{Average correct classification}\\ \hline
 \textbf{Concealed Weapon} & \cellcolor[gray]{.8}$91.675$ & $6.750$ & $1.575$ & \cellcolor[gray]{.8}$95.850$ & $2.525$ & $1.625$ & \cellcolor[gray]{.8}$98.100$ & $1.500$ & $0.400$ & $95.208$ \\ \hline
 \textbf{No Weapon} & $6.550$ & \cellcolor[gray]{.8}$92.475$ & $0.975$ & $1.350$ & \cellcolor[gray]{.8}$97.875$ & $0.775$ & $0.925$ & \cellcolor[gray]{.8}$97.675$ & $1.400$ & $96.008$  \\ \hline
 \textbf{Visible Weapon} & $0.575$ & $0.875$ & \cellcolor[gray]{.8}$98.550$ & $0.200$ & $0.775$ & \cellcolor[gray]{.8}$99.025$ & $0.525$ & $1.275$ & \cellcolor[gray]{.8}$98.200$ & $98.591$\\ \hline
& \textbf{Concealed Weapon} & \textbf{No Weapon} & \textbf{Visible Weapon} & \textbf{Concealed Weapon} & \textbf{No Weapon} & \textbf{Visible Weapon} & \textbf{Concealed Weapon} & \textbf{No Weapon} & \textbf{Visible Weapon}& \\ \hline
& \multicolumn{9}{c|}{\textbf{True Label}} & \\ \hline
\end{tabular}
\end{minipage}}
\end{table}

 Table \ref{tab:svmpred} contains the prediction output for event (a). The datasets captured for this event is represented by set $1(a)$, $2(a)$ and $3(a)$ in Table \ref{tab:traintestdet}. The prediction results are shown in terms of percentage of correct classification. In case 1, set $1(a)$ is used as test set when the classifier is trained using sets $2(a)$ and $3(a)$. Similarly in case 2, set $2(a)$ is used as test set when the classifier is trained using sets $1(a)$ and $3(a)$. Similarly in case 3, set $3(a)$ is used as test set when the classifier is trained using sets $1(a)$ and $2(a)$. Since the number of datasets used for training and testing in each of the cases is the same, the average correct classification is calculated by summing the correct classification for each set and dividing by three. The results show that the lowest classification occurs ($77.458\%$) in the scenario where no person is present in the room. Although the difference in prediction is not by a huge amount it can be explained by the fact that the reflections from a person's clothing might have similar features as the reflections from the curtains or the bed which the algorithm might be confusing as features of a person. Since there are only $40$ feature points per frame, all the details are not captured and some similar features can confuse the algorithm to give ambiguous results. 


 Table \ref{tab:svmpredweapon} contains the prediction output for event (b). The datasets captured for this event are represented by set $1(b)$, $2(b)$ and $3(b)$ in Table \ref{tab:traintestdet}. The number of correct classifications in percentage for each case is shown in Table \ref{tab:svmpredweapon}. In case 4, set $1(b)$ is used as test data when the classifier is trained using sets $2(b)$ and $3(b)$. Similarly in case 5, set $2(b)$ is used as test data when the classifier is trained using sets $1(b)$ and $3(b)$. Finally in case 6, set $3(b)$ is used as test data when the classifier is trained using sets $1(b)$ and $2(b)$. The number of feature points for this test scenario is $48$ per frame and that shows a significant improvement in the prediction percentage. The scenario with the minimum average correct prediction percentage is the scenario with a concealed weapon ($95.208\%$).



\section{Conclusion} \label{sec:conc}
 In this article we presented an application of GSM--CommSense towards detection of two different through-the-wall events. We have captured multiple sets of data for each event and presented a detailed analysis of each set using different machine learning algorithms. We show how data from different scenarios is clustered in a three dimensional PCA space and discuss the results. In addition to PCA clustering, we used a widely accepted supervised learning algorithm called SVM to check how accurate the predicted classification results are compared to the true values. The results are encouraging as a minimum average classification result obtained for detection of a person through a brick wall is $77.458\%$ with only $40$ points obtained from each GSM frame of $577$ \si{\micro\second} and $95.208\%$ in case of detection of weapon through a brick wall with $48$ points. The change in the performance result is dependent on multiple factors including the increase in the number of points per sample and the fact that the weapons (with aluminium foil) have significantly high radar cross section compared to a person. 
 
 The thickness and material of the wall is fixed as all the tests are performed in the same location. Changing the location changes many other environmental parameters that will cause in ambiguous categorization. Although if we train the system to detect across a certain wall it will always perform with the same amount of accuracy.
 

 The limitations for the current analysis is that the location of the person being detected across a wall is the same for all the cases. PCA used to observe clustering of the captured data shows only $52\%$ of the information which makes it difficult to visualize proper separation of the clusters. Thus SVM is used which takes the entire training dataset and transforms into $F-$dimensional space to characterize the test data. The limitation of SVM is that it can only be tested for events it has been trained to detect. Thus in real-time the outcome of this system will be very application specific. Distance of the objects from the wall was kept standard for this particular testing and more robust measurements needs to be taken varying more parameters.

\bibliography{central-bibliography-no3gppurl}
\end{document}